\documentclass[11pt, notitlepage, final]{article}
\newcommand{\ddx}{\frac{\partial}{\partial x}}
\newcommand{\ddxi}{\frac{\partial}{\partial x_i}}
\newcommand{\ddxj}{\frac{\partial}{\partial x_j}}
\newcommand{\ddy}{\frac{\partial}{\partial y}}

\newcommand{\dtdy}{\frac{\partial^2}{\partial y^2}}

\newtheorem{theorem}{Theorem}[section]

\usepackage{amsfonts , epsfig}

\begin{document}

\title{Turbulent boundary layer equations}
\author{A. Cheskidov\\ \\
\textit{Department of Mathematics}\\ 
\textit{Indiana University, Rawles Hall},
\textit{Bloomington, IN 47405, USA}\\
\texttt{http://php.indiana.edu/\~{}acheskid}\\
\texttt{acheskid@indiana.edu}}

\date{\today}
\maketitle

\begin{abstract}
We study a boundary layer problem for the Navier-Stokes-alpha model
obtaining a generalization of the Prandtl equations conjectured to
represent the averaged flow in a turbulent boundary layer.
We solve the equations for the semi-infinite plate, both
theoretically and numerically. The latter solutions agree with
some experimental data in the turbulent boundary
layer.
\end{abstract}

\section{Introduction}

Boundary-layer theory, first introduced by Ludwig Prandtl in 1904,
is now fundamental to many applications of fluid mechanics,
especially aerodynamics.

Consider the case of two-dimensional steady
incompressible viscous flow near a flat surface. Let $x$ be the
coordinate along the horizontal surface, $y$ be the
coordinate normal to the
surface, and $(u,v)$ be the corresponding velocity of the flow.
Near the wall $u$ is
significantly larger than $v$. Also, $u$ changes in $y$ much faster
than it does in the $x$ coordinate. Then, in the case of a zero
pressure gradient, the following Prandtl equations are used to
approximate the Navier-Stokes equations.
\begin{equation} \label {eq:Prandtl}
\left\{
\begin{array}{l}
u \ddx u + v \ddy u = \nu \dtdy u\\
\ddx u + \ddy v = 0,
\end{array}
\right.
\end{equation}
where $\nu$ is a kinematic viscosity; density is chosen to be
identically one.

A classical example of the boundary layer is a flow near the
semi-infinite plate $\{(x,y):x\geq 0, y=0\}$.  
In 1908, Blasius discovered that in this case there exists
a similarity variable $\xi = \frac{y}{\sqrt{x}}$.
Equations (\ref{eq:Prandtl}) can be reduced to an ordinary differential
equation
\begin{equation} \label{BlasiusODE}
h'''+\frac{1}{2}hh''=0
\end{equation}
with $h(0)=h'(0)=0$, and
$h'(\xi) \rightarrow 1$ as $\xi \rightarrow \infty$. 
Then
\[
u(x,y) = u_e h'\left( \frac{y}{\sqrt{l_*x}}\right), \ \
v(x,y) = \frac{u_e}{\sqrt{R_x}} h'\left(\frac{y}{\sqrt{l_*x}} \right)
\]
are solutions of (\ref{eq:Prandtl}) and they match experimental data
in the laminar boundary layer. Here $u_e$ is the  horizontal velocity
component of the external flow, $l_*=\frac{\nu}{u_0}$, and
$R_x=\frac{x}{l_*}$.

For high local Reynolds numbers $R_x$ the flow becomes turbulent and analogs
of the equations (\ref{eq:Prandtl}) and (\ref{BlasiusODE}) were not
known. In this paper we use  Navier-Stokes-alpha
model of fluid turbulence also known as viscous Camassa-Holm equations
to study averaged flow, and obtain turbulent boundary layer
equations that generalize (\ref{eq:Prandtl}) and (\ref{BlasiusODE}).
Our reduction to an ordinary differential equation (see (\ref{eq:1d}))
also uses Blasius's similarity variable.

Alpha-model was first introduced in \cite{HMR-98a} as a generalization
to $n$ dimensions of the one-dimensional Camassa-Holm equation that
describes shallow water waves. Later, NS-$\alpha$
model was proposed as a closure approximation for the Reynolds equations,
and its solutions were compared with empirical data for turbulent
flows in channels and pipes
\cite{Chen-etal-PRL[1998]}-\cite{Chen-etal-PhysD[1999]}.

\section{Derivation}
\label{sec:derivation}

We study two-dimensional steady incompressible viscous flow near a
flat surface. Let $x$ be the coordinate
along the surface, $y$ be the coordinate normal to the surface.
Denote also $(u,v)$ to be the velocity of the flow.

2-D Navier-Stokes-$\alpha$ model is used to study a boundary layer
flow.
\begin{equation} \label{CH:eq}
\left\{
\begin{array}{l}
\frac{\partial}{\partial t} \mathbf{v} +
(\mathbf{u} \cdot \nabla)\mathbf{v}+v_j \nabla u_j = 
\nu\Delta \mathbf{v} - \nabla q \\
\nabla \cdot \mathbf{u} = 0,
\end{array}
\right.
\end{equation}
where $\mathbf{u}=(u,v)$, and
$\mathbf{v}=(\gamma,\tau)$ is a momentum defined in the following way:
\[
\mathbf{v}=\mathbf{u}-\ddxi \left(\alpha^2 \delta_{ij}
 \ddxj \mathbf{u}\right).
\]
Non-slip boundary conditions are used: $\mathbf{u}|_{y=0}=0$. The other
boundary conditions are going to be determined later.

Fix $l$ on the $x$-axis and define $\epsilon(l)$ to be
$\epsilon:=1/\sqrt{R_l}=\sqrt{\nu/(u_e l)}$,
where $\nu$ is viscosity and $u_e$ is the horizontal velocity component of the external
flow.

We change variables $x_1=\frac{1}{l}x$,
$y_1=\frac{1}{\epsilon l}y$, $u_1=\frac{1}{u_e}u$,
$v_1=\frac{1}{\epsilon u_e}v$, $q_1=\frac{1}{u_e^2}q$. 
$\alpha_1=\frac{\alpha}{\epsilon l}$.
In addition, assume that $\alpha_1$ doesn't depend on $y$ variable.
After neglecting terms with high powers of $\epsilon$, dropping
subscripts and denoting
\[
w=\left(1-\alpha^2\dtdy\right)u,
\]
we derive the following turbulent boundary layer equations that
generalize Prandtl equations:
\begin{equation} \label{eq:gPrandtl}
\left\{
\begin{array}{l}
u\ddx w + v\ddy w + w\ddx u = \dtdy w - \ddx q \\
w\ddy u = -\ddy q \\
\ddx u + \ddy v =0.
\end{array}
\right.
\end{equation}
Note that derivative of $q$ is not zero in $y$-direction. Therefore,
we introduce
$
Q(x,y):=q+\frac{1}{2}u^2-\frac{1}{2}\alpha^2\left(\ddy u\right)^2,
$
and obtain the following equations:
\[ 
\left\{
\begin{array}{l}
u\ddx w + v\ddy w + \alpha^2\left(\frac{\partial u}{\partial y}
\frac{\partial^2 u}{\partial x \partial y} - \frac{\partial u}{\partial x}
\frac{\partial^2 u}{\partial y^2}\right) + \frac{1}{2}\ddx \alpha^2 \cdot 
\left(\ddy u\right)^2
= \dtdy w - \ddx Q \\
\ddy Q = 0 \\
\ddx u + \ddy v =0.
\end{array}
\right.
\]

Consider now a case of a two-dimensional
steady incompressible viscous flow near the semi-infinite plate
$\{(x,y):x\geq 0, y=0\}$.
The following assumptions are made in this case:
\begin{itemize}
\item
$\alpha=\sqrt{x}\beta$
\item
Zero pressure gradient, i.e. $\ddx Q=0$.
\end{itemize}
In addition, we will study the solutions $(u_\infty, v_\infty)$ of
(\ref{eq:gPrandtl}), which on
some adequate interval $x_1\leq x\leq x_2$ are of the form
\[
u_{\infty}=f(\xi), \ v_{\infty}=\frac{1}{\sqrt{x}}g(\xi), \ \xi=\frac{y}{\sqrt{x}}.\]

Let
$
h(\xi)=\int_0^\xi f(\eta)d\eta.
$
Then $g=\frac{1}{2}\xi h'-\frac{1}{2}h$, and we have the following equation for
$h$:
 \begin{equation} \label{eq:1d}
h'''+\frac{1}{2}hh''-\beta^2\left(h'''''+\frac{1}{2}hh''''\right)=0.
\end{equation}
The boundary conditions are $h(0)=h'(0) =0$, $h''(0)>0$, and
$h'(\xi) \rightarrow 1$ as $\xi \rightarrow \infty$.
Note that if $h(\xi)$ is a solution of (\ref{eq:1d}),
then $\hat{h}(x):=\beta h(\beta x)$ is a solution of
\begin{equation} \label{eq:inx}
h'''+\frac{1}{2}hh''-\left( h'''''+\frac{1}{2}hh''''\right)=0.
\end{equation}
This equation can be also written as
$
m'''+\frac{1}{2}hm''=0,
$
where $m=h-h''$.
The following theorem is proven for this equation in \cite{C-2001}.
\begin{theorem}
Given any $a>0$, $b$, there exists $c(a,b)$ such that
$h'(\xi)\rightarrow const. \geq 0$ as $\xi \rightarrow \infty$, where
$h$ is a solution of (\ref{eq:inx}) with $h(0)=h'(0)=0$, $h''(0)=a$,
$h'''(0)=b$, $h''''(0)=c$.
\end{theorem}

\section{Comparison with Experimental Data}
It is common to use
\[
y^+=\frac{u_\tau y}{\nu}, \ \ u^+=\frac{u}{u_\tau}
\]
in the turbulent boundary layer, where
$
u_\tau= \sqrt{\nu\frac{\partial u}{\partial y}\Big|_{y=0}}.
$
Fix $x$ on the horizontal axis and denote
$l_*=\frac{\nu}{u_e}$, $R_x=\frac{x}{l_*}$.
As argued in the section \ref{sec:derivation},
\[
u=u_e h'\left(\frac{y}{\sqrt{l_*x}}\right)
\]
represents a horizontal component of the averaged velocity for
some $h$ satisfying (\ref{eq:1d})
with $h(0)=h'(0)=0$, $h''(0)=a>0$, $h'''(0)=b$, 
and $h'(\xi) \rightarrow 1$ as $\xi \rightarrow \infty$.

For any such $h$, $\hat{h}(\xi)=\beta h\left(\beta \xi\right)$
satisfies (\ref{eq:inx}).
Then $\beta^2 = \lim_{\xi \rightarrow \infty} \hat{h}'(\xi)$. In addition,
\[
u^+=\frac{R_x^{1/4}}{\sqrt{a}\beta^2 }\hat{h}' \left(
\frac{y^+ }{\sqrt{a}\beta R_x^{1/4}}  \right).
\]
Given $c_f$, a skin-friction coefficient and $R_\theta$, a Reynolds
number based on momentum thickness we find $a$, $b$,
and $R_x$ so that the following conditions hold:
\begin{enumerate}
\item
$c_f=2/ \left(\inf_y u^+\right)^2.$
\item
$
R_\theta = \frac{1}{\nu}\int_0^\delta u\left(1-\frac{u}{u_e}\right) dy.
$
\item
Von Karman log law for the middle inflection point in logarithmic coordinates.
\end{enumerate}

\begin{figure}[e]
\center
  \epsfxsize=12cm
  \epsfysize=4.5cm
  \epsfbox {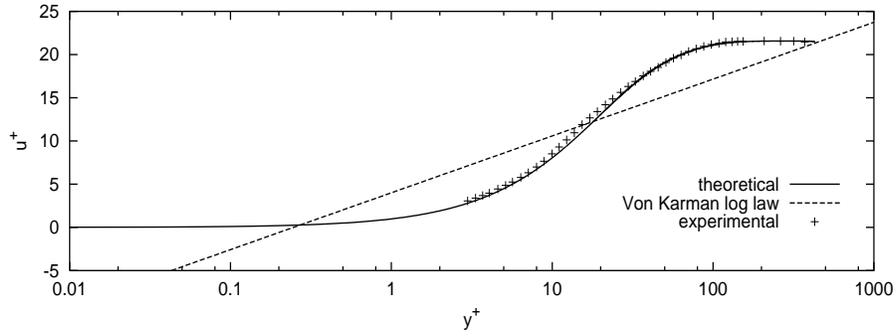}
\caption{$c_f=4.32, \  R_\theta =  265, \ \beta=2.22$}
\label{gr:0}
\end{figure}

\begin{figure}
\center
  \epsfxsize=12cm
  \epsfysize=4.5cm
  \epsfbox {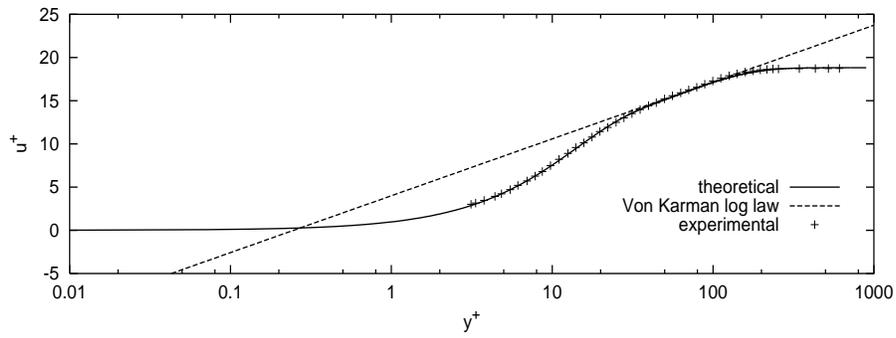}
\caption{$c_f=5.69, \  R_\theta =  396, \ \beta=6.63$}
\label{gr:3}
\end{figure}

\begin{figure}
\center
  \epsfxsize=12cm
  \epsfysize=4.5cm
  \epsfbox {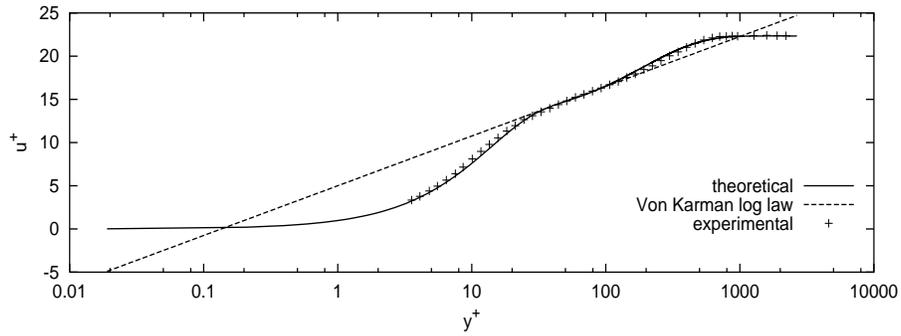}
\caption{$c_f=4.01, \  R_\theta =  1436, \ \beta=19.8$}
\label{gr:6}
\end{figure}

A family of curves $\{ u \} _{c_f, R_\theta}$ was
compared with experimental data of Rolls-Royce applied
science laboratory, ERCOFTAC t3b test case (see Fig. \ref{gr:0},
\ref{gr:3}, and \ref{gr:6}). 
Comparison shows that the case
$a+\beta b>0$ corresponds to a laminar
region of a boundary layer, $a+\beta b<0$ corresponds to a
turbulent region of a boundary layer. The case $a+\beta b=0$
corresponds to a transition point.

\textbf{Acknowledgment.}
This work was supported
in part by NSF grants DMS-9706903, DMS-0074460.


\end{document}